\documentclass[aps,prb,twocolumn, showpacs]{revtex4}

\usepackage{epsfig}
\usepackage{graphicx}

\def\be{\begin{equation}}
\def\ee{\end{equation}}
\def\bea{\begin{eqnarray}}
\def\eea{\end{eqnarray}}

\begin{document}
\draft

\title{Efficient electron spin manipulation in a quantum well by an in-plane electric field}
\author{E. I. Rashba$^{1}$\cite{Rashba*} and Al. L. Efros$^2$}
\affiliation{$^1$Department of Physics, SUNY at Buffalo, Buffalo, New York 14260, USA\\
$^2$Naval Research Laboratory, Washington, DC 20375, USA}

\begin{abstract}
Electron spins in a semiconductor quantum well couple to an electric field {\it via} spin-orbit interaction.  We show that the standard spin-orbit coupling mechanisms can provide extraordinary efficient electron spin manipulation by an in-plane ac electric field.
\end{abstract}

\maketitle

\narrowtext

Efficient manipulation of electron spins by an external ac field is one of the central problems of semiconductor spintronics \cite{reviews}, quantum computing and information processing. \cite{LDV} The original proposals of spin manipulation were based on using a time dependent magnetic field $\tilde{\mbox{\boldmath$B$}}(t)$. However, there is a growing understanding of the advantages of  spin manipulation by a time-dependent electric field $\tilde{\mbox{\boldmath$E$}}(t)$ that couples to the electron spin through different mechanisms of spin-orbit (SO) interaction.\cite{RS91} Recently Kato {\it et al.} successfully manipulated electron spins in a parabolic AlGaAs quantum well (QW) by a gigahertz bias applied to a single gate; this produced a field $\tilde{\mbox{\boldmath$E$}}(t)$ perpendicular to the well.\cite{Kato} The structure was engineered to make the electron Land\'{e} tensor $\hat g$ small ($|g|\alt 0.1$), highly anisotropic, and position-dependent, and to achieve in this way a rather strong SO coupling originating from the position-dependence of the Zeeman energy. The present authors have shown\cite{RE03}  that for large $g$-factors typical of narrow-gap A$_3$B$_5$ semiconductors such a field $\tilde{\mbox{\boldmath$E$}}(t)$ can provide efficient electrical operation {\it via} the standard mechanisms of SO coupling (Dresselhaus\cite{D55} and Rashba\cite{BR84}) but only for relatively wide QWs. Indeed, the coupling of 2D electron spins to a perpendicular field $\tilde{\mbox{\boldmath$E$}}(t)$ develops due to a deviation from the strict two-dimensional (2D) limit and is proportional to $w/\lambda$ (for $w/\lambda\ll 1$), where $w$ and $\lambda$ are the confinement and magnetic lengths, respectively. Using a tilted magnetic field \mbox{\boldmath$B$} is critical for the mechanisms of Refs.~\onlinecite{Kato} and \onlinecite{RE03}.

In this paper we show that using an in-plane electric field $\tilde{\mbox{\boldmath$E$}}(t)$ results in a significant increase in the coupling of this field to electron spins. Moreover, the coupling exists in the strict 2D limit (and usually increases with decreasing $w$) and for arbitrary orientations of \mbox{\boldmath$B$}, including a perpendicular-to-plane \mbox{\boldmath$B$}. However, using a tilted field \mbox{\boldmath$B$} allows distinguishing the contributions coming from the competing  mechanisms of SO coupling. Of course, producing an in-plane field $\tilde{\mbox{\boldmath$E$}}(t)$ needs designing a proper coupling to the radio frequency or microwave source; this problem can be solved.

We consider a 2D electron gas in a A$_3$B$_5$ crystal confined in a (0,0,1) QW subject to a tilted magnetic field $\mbox{\boldmath$B$}$. The 2D kinetic momentum of electrons $\hat{\mbox{\boldmath$k$}}=-i\mbox{\boldmath$\nabla$}+e\mbox{\boldmath$A$}/\hbar c$ depends only on the normal component of \mbox{\boldmath$B$}, $B_z=B\cos\theta$, hence $\mbox{\boldmath$A$}(\mbox{\boldmath$r$})=(B_z/2)(-y,x,0)$. The operators $\hat{k}_x$ and $\hat{k}_y$ obey the usual commutation relations $[\hat{k}_x,\hat{k}_y]_-=-i/\lambda^2$, where $\lambda=(c\hbar/eB_z)^{1/2}$; we assume that $B_z>0$. The total Hamiltonian $\hat{H}=\hat{H}_0+\hat{H}_Z+\hat{H}_{\rm so}+\hat{H}_{\rm int}$ includes the orbital term $\hat{H}_0=\hbar^2\hat{\mbox{\boldmath$k$}}^2/2m$, where $m$ is the electron effective mass, the Zeeman term $\hat{H}_Z=g\mu_B({\mbox{\boldmath$\sigma$}}\cdot\mbox{\boldmath$B$})/2$, $\mu_B$ being the Bohr magneton, and the coupling $\hat{H}_{\rm int}(t)=e(\mbox{\boldmath$r$}\cdot{\tilde{\mbox{\boldmath$E$}}}(t))$ to an in-plane ac electric field $\tilde{\mbox{\boldmath$E$}}(t)$. The SO interaction $\hat{H}_{\rm so}=\hat{H}_D+\hat{H}_R$ includes the Dresselhaus and Rashba terms that in the principal crystal axes are
\begin{equation}
\hat{H}_D=\alpha_D(\sigma_x\hat{k}_x-\sigma_y\hat{k}_y),~~
\hat{H}_R=\alpha_R(\sigma_x\hat{k}_y-\sigma_y\hat{k}_x).
\label{eq1}
\end{equation}

It is convenient to change from the set of operators $(x,y,-i\partial_x,-i\partial_y)$ to the kinetic momenta $(\hat{k}_x,\hat{k}_y)$ and the coordinates of the center of the orbit, $(\hat{x}_0,\hat{y}_0)$.\cite{JL49} The latter ones commute with $(\hat{k}_x,\hat{k}_y)$ and with $H_0$ and obey the commutation relations $[\hat{x}_0,\hat{y}_0]_-=i\lambda^2$. As usual, instead of $(\hat{k}_x,\hat{k}_y)$ the Bose-operators $\hat{a}=\lambda(\hat{k}_x-i\hat{k}_y)/\sqrt{2}$ and $\hat{a}^+=\lambda(\hat{k}_x+i\hat{k}_y)/\sqrt{2}$ can be used. After these transformations, we come to the convenient expressions for the coordinates that appear in the operator $H_{\rm int}(t)$ 
\begin{equation}
x=\hat{x}_0-i\lambda(\hat{a}^+-\hat{a})/\sqrt{2},~~ y=\hat{y}_0-\lambda(a^++a)/\sqrt{2}.
\label{eq2}
\end{equation}
Without the SO interaction, the energy spectrum of 2D electrons is described by two sets of Landau levels $E_\sigma(n)=\hbar\omega_c(\theta)(n+1/2)+\hbar\omega_s\sigma$ for two spin projections on the \mbox{\boldmath$B$} direction, $\sigma=\pm1/2$. Here $\omega_c(\theta)=eB_z/\hbar c=\omega_\perp\cos\theta$ and $\omega_s=g\mu_BB/\hbar$ are the cyclotron and spin flip frequencies, respectively, and $n\geq0$.  The energy spectrum of a 2D system with a single term in $\hat{H}_{\rm so}$, either $\hat{H}_D$ or $\hat{H}_R$, can be found analytically but only for a perpendicular field \mbox{\boldmath$B$}. The problem with both terms in $H_{\rm so}$ cannot be solved analytically. We consider in what follows the limit of a strong magnetic field (that is of principal physical interest and allows solving the problem for an arbitrary direction of \mbox{\boldmath$B$}) and assume that the frequencies $\omega_c$ and $\omega_s$ are large compared with the SO coupling, $\hat{H}_{\rm so}\ll\hbar\omega_c,\hbar\omega_s$. Then the SO term can be eliminated, in the first order in $\hat{H}_{\rm so}$, by a canonical transformation $\exp(\hat{T})$.\cite{RS91} The operator $\hat{T}$ is non-diagonal in the orbital quantum number $n$ and its matrix elements are
\begin{equation}
\langle n^\prime,\sigma^\prime\vert \hat{T} \vert n,\sigma\rangle=
\langle n^\prime,\sigma^\prime\vert \hat{H}_{\rm so}\vert n,\sigma\rangle/[E_{\sigma^\prime}(n^\prime)-E_\sigma(n)].
\label{eq3}
\end{equation}
After the transformation, the time independent part of $\hat{H}$ conserves  the spin projection on the magnetic field. 

The operator $\mbox{\boldmath$r$}=(x,y)$ is diagonal in the spin indices, and $\hat{H}_{\rm int}(t)$ drives spin-flip transitions due to the level mixing produced by $\hat{H}_{\rm so}$. After the $T$-transformation,  
\mbox{\boldmath$r$} acquires an anomalous part $\mbox{\boldmath$r$}_{\rm so}=[\hat{T},\mbox{\boldmath$r$}]_-$ that drives spin transitions. The matrix elements 
of $\mbox{\boldmath$r$}_{\rm so}$ 
diagonal in $n$ are
\begin{eqnarray}
\langle n\uparrow\vert&\mbox{\boldmath$r$}_{\rm so}&\vert n\downarrow\rangle=
-\bigl\{\omega_c\langle n\uparrow\vert[\hat{H}_{\rm so},\mbox{\boldmath$r$}]_+\vert n\downarrow\rangle\nonumber\\
&+&
\omega_s\langle n\uparrow\vert[\hat{H}_{\rm so},\mbox{\boldmath$r$}]_-\vert n\downarrow\rangle\bigr\}/\hbar(\omega_c^2-\omega_s^2);
\label{eq4}
\end{eqnarray}
the subscripts $+$ and $-$ designate anticommutators and commutators, respectively. Because $\hat{H}_{\rm so}$ includes only the operators $(\hat{a},\hat{a}^+)$, the operators $(\hat{x}_0,\hat{y}_0)$ appearing in Eq.~(\ref{eq2}) make no contribution to the diagonal in $n$ matrix elements of Eq.~(\ref{eq4}). These matrix elements can be calculated explicitly for $\hat{H}_D$ and $\hat{H}_R$.

In the quantum limit, when only the lowest Landau level $n=0$ is populated, for a magnetic field $\mbox{\boldmath$B$}=B(\sin\theta\cos\varphi, \sin\theta\sin\varphi, \cos\theta)$ and an in-plane electric field $\tilde{\mbox{\boldmath$E$}}(t)$ polarized at an angle $\psi$ to the $x$ axis, $\tilde{\mbox{\boldmath$E$}}(t)=\tilde{E}(t)(\cos\psi, \sin\psi, 0)$, the matrix elements $l_{D,R}^\parallel=\langle 0\uparrow\vert x_{\rm so}\cos\psi+y_{\rm so}\sin\psi\vert 0\downarrow\rangle_{D,R}$ of the electric dipole spin resonance (EDSR) are
\widetext 
\begin{eqnarray}
l_D^\parallel&=&{{\alpha_D}\over{\hbar(\omega_c^2-\omega_s^2)}}
[(\omega_c\cos\theta-\omega_s)\sin(\varphi+\psi)-i(\omega_c-\omega_s\cos\theta)\cos(\varphi+\psi)],
\label{eq5}\\
l_R^\parallel&=&-~
{{\alpha_R}\over{\hbar(\omega_c^2-\omega_s^2)}}
[(\omega_c\cos\theta+\omega_s)\cos(\varphi-\psi)+i(\omega_c+\omega_s\cos\theta)\sin(\varphi-\psi)].
\label{eq6}
\end{eqnarray}
\endwidetext 
\noindent
We assume that $\cos\theta>0$. When $\cos\theta<0$, one should also change the sign of $\omega_c$, hence, matrix elements are even with respect to the reflections in the $(x,y)$ plane. For $n\neq 0$, the anticommutator in Eq.~(\ref{eq4}) acquires a factor $(2n+1)$ while the commutator does not depend on $n$. Therefore, Eqs.~(\ref{eq5}) and (\ref{eq6}) can be generalized by substituting $\omega_c\rightarrow(2n+1)\omega_c,~ \omega_s\rightarrow\omega_s$ in the numerators.
\begin{figure}[th]
\vskip 0.2truecm
\begin{center}
\epsfig{file=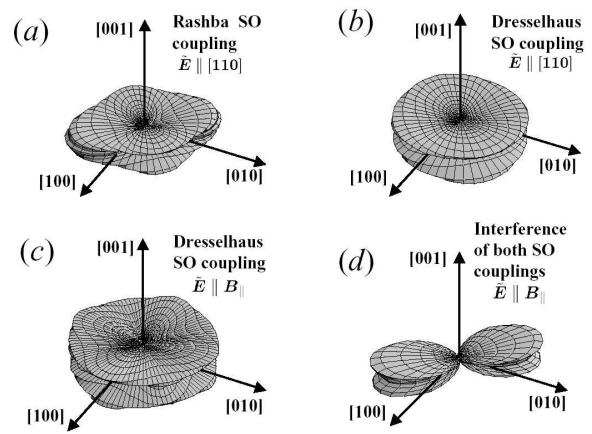, angle=-0, width=0.45\textwidth}
\end{center}
\vskip-0.5truecm
\caption{ Angular dependence of the EDSR intensity $I(\theta,\varphi,\psi)\propto |\l_R^\parallel+ l_D^\parallel|^2$ for a (0,0,1) QW calculated from Eqs.~(\ref{eq5}) and (\ref{eq6}) for: (a) --  $\alpha_D=0$ and $\psi=\pi/4$; (b) -- $\alpha_R=0$ and $\psi=\pi/4$; (c) -- $\alpha_R=0$ and $\psi =\varphi$; and (d) -- $\alpha_R=\alpha_D$ and $\psi=\varphi$; arbitrary units.  The ratio of the frequencies $\omega_s/\omega_\perp=-0.17$ as in InAs. The pole in $I(\theta,\varphi,\psi)$ at $\omega_c^2(\theta)=\omega_s^2$ was cut off by adding an imaginary part $i\Gamma=0.1 i\omega_\perp$ to $\omega_c(\theta)$. For large $\theta$ values, in the vicinity of the pole and close to $\pi/2$, the figure provides only qualitative patterns of $I(\theta,\varphi,\psi)$ because of its high sensitivity to the phenomenological line width $\Gamma$ and the restrictions imposed by the condition $\hbar\omega_c\gg \hat{H}_{\rm so}$.}
\end{figure}
\vskip-0.0truecm 
Let us first discuss the angular dependence of these matrix elements. The Hamiltonian of the Rashba interaction $\hat{H}_R$ is an invariant of the group $\mbox{\boldmath$C$}_{\infty v}$ of the continuous rotations about the $z$ axis. Therefore, $l_R^\parallel$ is isotropic with respect to the joint rotations of $\tilde{\mbox{\boldmath$E$}}$ and \mbox{\boldmath$B$} and depends only on the difference $(\varphi-\psi)$. However, this dependence is rather strong and the sign of the azimuthal anisotropy depends on $\theta$, Fig.~1a. For the Dresselhaus interaction, $l_D^\parallel$ is anisotropic, Figs.~1b and 1c. When $\varphi\rightarrow\varphi+\pi/2$ and $\psi\rightarrow\psi+\pi/2$, $l_D^\parallel$ changes sign. As a result, the intensity of EDSR that is proportional to ${l_D^\parallel}^2$ shows a four-fold symmetry with respect to the joint rotations of \mbox{\boldmath$B$} and $\tilde{\mbox{\boldmath$E$}}$ about the $z$ axis, Fig.~1c. Generally, the intensity is proportional to $(l_D^\parallel+l_R^\parallel)^2$, hence, both contributions interfere and the intensity shows only the two-fold symmetry in accordance with the symmetry group $\mbox{\boldmath$C$}_{2v}$ of the Hamiltonian $\hat{H}_{\rm so}$. When $\hat{H}_D$ and $\hat{H}_R$ are of a comparable magnitude, as  was found for GaAs QWs,\cite{GaAs1,GaAs2,GaAs3,GaAs4} the effect of the interference is very strong, see Fig.~1d. Therefore, {\it the azimuthal dependence of the intensity of EDSR is a powerful tool for measuring the ratio of the coupling constants, $\alpha_R/\alpha_D$}. 

It is seen from Eqs.~(\ref{eq5}) and (\ref{eq6}) that the EDSR driven by an in-plane field $\tilde{\mbox{\boldmath$E$}}$ should be seen for any direction of \mbox{\boldmath$B$} including perpendicular to the QW plane. However, using a tilted field \mbox{\boldmath$B$} provides special advantages. First, it allows distinguishing the contributions of the different SO coupling mechanisms. Second, because of the poles in the denominators at $\omega_c(\theta)\approx\omega_s$, it can allow strong increase in the EDSR intensity. A similar resonance in the EDSR intensity when the frequency of an electric dipole transition approaches $\omega_s$ is known in the spectroscopy of local centers.\cite{RS91,Dobr84} Near the pole, where the denominator of Eq.~(\ref{eq4}) vanishes, perturbation theory in the interaction $\hat{H}_{\rm so}$ fails and the sharpness of the resonance is cut by the avoided level crossing and by the level width.  

Let us  estimate now the magnitudes of the matrix elements of Eqs.~(\ref{eq5}) and (\ref{eq6}) for in-plane EDSR, $\tilde{\mbox{\boldmath$E$}}\perp{\hat{\mbox{\boldmath$z$}}}$, and compare them to the matrix elements for the different mechanisms of spin-flip transitions. By the order of magnitude, $l_D^\parallel\approx\alpha_D/\hbar\omega_c$ and $l_R^\parallel\approx\alpha_R/\hbar\omega_c$ for $\mbox{\boldmath$B$}\parallel{\hat{\mbox{\boldmath$z$}}}$. With typical values of $\alpha_D\approx\alpha_R\approx 10^{-9}$ eV cm, $m\approx 0.05m_0$, and $B\approx 1$ T, we get $l_D^\parallel\approx l_R^\parallel\approx 10^{-5}$ cm. A similar length for the electron paramagnetic resonance (EPR) is $l_{\rm EPR}\approx |g|\lambdabar_C/4\approx 10^{-10}$ cm for $|g|\approx 10$; $\lambdabar_C=\hbar/m_0c$ being the Compton length. Hence, $l_D^\parallel, l_R^\parallel\gg l_{\rm EPR}$, and {\it EDSR strongly dominates over EPR}.

In the perpendicular geometry, $\tilde{\mbox{\boldmath$E$}}\parallel \hat{\mbox{\boldmath$z$}}$, the characteristic length for the Rashba interaction is $l_R^\perp\approx\alpha_R\omega_s/\hbar\omega_0^2$, where $\omega_0\approx \hbar/mw^2$ is the confinement frequency. The factor $\omega_s$ appears as a result of the electron confinement in the electric field direction; its presence is required by the Kramers theorem.\cite{RS91} The ratio of the characteristic lengths $l_R^\perp/l_R^\parallel\approx\omega_c\omega_s/\omega_0^2\approx(mg/2m_0)(w/\lambda)^4$. Therefore, under strong confinement conditions, $w\ll\lambda$, {\it EDSR in the in-plane geometry is much stronger than in the perpendicular geometry} because in the latter geometry it develops due to the deviation from the 2D regime.\cite{RE03}

For the Dresselhaus interaction, it is instructive to compare the EDSR intensity in the in-plane geometry to its intensity in 3D.\cite{RS61} Usually the 3D Dresselhaus length is about $l_D^{3D}\approx\delta/\hbar\omega_c\lambda^2$. Here $\delta$ is the constant of the bulk inversion asymmetry of A$_3$B$_5$ compounds and is related to $\alpha_D$ as $\alpha_D\approx\delta/w^2$.\cite{coef} The ratio of the characteristic lengths is about $l_D^\parallel/l_D^{3D}\approx(\lambda/w)^2$. There is a special 3D geometry, $\tilde{\mbox{\boldmath$E$}}\parallel\mbox{\boldmath$B$}$, when $l_D^{3D}$ is especially large, $l_D^{3D}\approx\delta/\hbar\omega_s\lambda^2$; indeed, $\omega_s/\omega_c$ is usually numerically small. However, in a similar 2D geometry, $\tilde{\mbox{\boldmath$E$}}\parallel\mbox{\boldmath$B$}\perp{\hat{\mbox{\boldmath$z$}}}$, the length $l_D^\parallel$ is also large, $l_D^\parallel\approx\alpha_D/\hbar\omega_s$, as follows from Eq.~(\ref{eq5}) for $\cos\theta=0$ and $\omega_c=0$. Hence, again $l_D^\parallel/l_D^{3D}\approx(\lambda/w)^2$. Therefore, under the conditions of a strong confinement, $w\ll \lambda$, {\it in-plane EDSR in a QW is stronger than in 3D}. Absence of the potential confinement in the direction of $\tilde{\mbox{\boldmath$E$}}$ and a strong confinement in $z$ direction are highly advantageous for strong EDSR.

The most important quantity characterizing the spin operation efficiency by a resonant electric field ${\tilde E}(t)$ is the Rabi frequency $\Omega_R=e{\tilde E}l/\hbar$. With $l\approx \l_R^\parallel, l_D^\parallel\approx 10^{-5}$ cm as estimated above, we find that $\Omega_R\approx 10^{10} {\rm s}^{-1}$ in an electric field as small as only about ${\tilde E}\approx 0.6$\,V/cm. This estimate shows that {\it the electron spin manipulation by an in-plane electric field should be highly efficient}.

Electron heating by the electric field and spin relaxation can hamper electrical operation of electron spins. However, electron heating is suppressed by a strong magnetic field because, according to the Drude formula, it decreases with $B$ as $(\omega_c\tau_p)^{-2}$, $\tau_p$ being the momentum relaxation time. The spin relaxation is universal, i.e., it does not depend on the specific mechanism of spin operation. The above theory has been developed for the 2D limit but the qualitative conclusions related to the high efficiency of in-plane operation are valid also when $\omega_0\approx\omega_c$.

In conclusion, we have demonstrated an extraordinary high efficiency for the electron spin operation in quantum wells by an in-plane ac electric field.  Spin coupling to an in-plane ac electric field is many orders of magnitude stronger than to an ac magnetic field, and is also much stronger that the coupling to a perpendicular-to-well ac electric field. 

We are grateful to T. A. Kennedy for useful suggestions. E.I.R. \& Al.L.E. acknowledge the financial support from DARPA/SPINS by the ONR Grant N000140010819 and from DARPA/QuIST and ONR, respectively.


\begin{thebibliography}{99}
\bibitem{coef} A numerical coefficient in this formula depends on the shape of the confining potential. We disregard it similarly to the numerical coefficients in the different estimates.
\end{thebibliography}
\end{document}